# Near field radiative heat transfer between two nonlocal dielectrics

F. Singer, Y. Ezzahri and K. Joulain*
Institut Pprime, Université de Poitiers-CNRS-ENSMA
2, Rue Pierre Brousse, Bâtiment B25, TSA 41105
86073 Poitiers Cedex 9, France.
* : karl.joulain@univ-poitiers.fr

## Abstract

We explore in the present work the near-field radiative heat transfer between two semi-infinite parallel nonlocal dielectric planes by means of fluctuational electrodynamics. We use atheory for the nonlocal dielectric permittivityfunction proposed byHalevi and Fuchs. This theory has the advantage to includedifferent models performed in the literature. According to this theory, the nonlocal dielectric function is described by a Lorenz-Drude like single oscillator model, in which the spatial dispersion effects are represented by an additional term depending on the square of the total wavevector $k$. The theory takes into account the scattering of the electromagneticexcitation at the surface of the dielectric material, which leads to the need of additional boundary conditions in order to solve Maxwell's equations and treat the electromagnetic transmission problem. The additional boundary conditions appear as additional surface scattering parameters in the expressions of the surface impedances. It is shown that the nonlocal modeling deviates from the classical $1/d^2$ law in the nanometerrangeat distances still larger than the ones where quantum effects are expected to come into play.

# I. Introduction

In the last two decades, a growing theoretical and experimental research has been devoted to the study of radiative heat transfer at distances much smaller than the typical wavelength of thermal radiation [1-7]. This so-called near field radiative heat transfer follows physical laws that are different from the ones governing classical radiative heat transfer i.e. the laws of geometrical optics. At subwavelength distances, the wave behavior of light has to be considered and phenomena such as tunneling or interferences control radiative heat transfer. These phenomena completely change the usual behavior of radiative heat transfer which is classically seen as a broadband signal limited in intensity to the exchanges between blackbodies. In the near-field, radiative heat transfer which is ruled by the density of electromagnetic states can be strongly changed due the presence of additional modes at certain frequencies: radiative heat transfer can surpass classical radiation due to the presence of modes close to the surface able to tunnel between heated bodies [8-10]. These new features have open the way to the search of very promising energetics applications such as near field thermophotovoltaics.Indeed, the control of the near-field thermal radiation could lead to a quasi-monochromatic transfer enhanced by several orders of magnitude from the far field values and potentially leading to high conversion ratios [11-17]. Other applications such as cooling [18], nanolithography [19,20] or subwavelength source [21] are concerned with these physicslaws changes at subwavelength scales.

Experimental research has confirmed near field radiative heat transfer theoretical predictions. The thermal density of energy is much higher in the near field in comparison to the far field, which is due to the presence of surface waves [22], whereas near field radiative heat transfer between bodies at different temperatures is increased as well as in tip-surface geometry [23-25] or in plane parallel geometry [26-30]. Moreover, the change in thermal radiation spectral content has also been observed in the near field [31-33], where a quasi-monochromatic spectral behavior has been reported above SiC and SiO2.

In the work presented here, we will focus on the radiative heat transfer behavior between two heated semi-infinite parallel dielectric solid planes at small distance of separation *d*. In past theoretical studies, it has been shown that near field radiative heat transfer follows a *1/d²* law as long as the separation distance is of the order of few hundreds of nanometers [2-4,9]. Metals follow a quite different behavior due to the presence of magnetic effects which are surpassed by the *1/d²* law only at distances below the angstrom range [2,34]. At such low separation distances, fluctuational electrodynamics has to be questioned, in particular the fact that the material optical response is still local. Moreover, the fact that radiative heat transfer is the dominant heat transfer mode has also to be questioned. Of course, when the separation distances are going to be around the typical atomic distances in matter, quantum effects could appear especially for metals where electrons are the dominant heat carriers [35-37] but also in dielectrics for which quantum effects influence has been recently proved with molecular dynamics [38]. At these interatomic separation distances, transition to a regime where thermal conduction dominates occurs. However, it still remains an open question about whether corrections due to the nonlocal optical response of the material appear at distances larger than the one where quantum effects appear and at what distances these nonlocal effects prevail.

To the best of our knowledge, no nonlocal correction to the radiative near field heat transfer has been addressed in the past in the case of dielectrics apart from a very phenomenological description [39]. In the case of metals however, an important and complete work has been performed by Chapuis et al. [34] using the Lindhard-Mermin nonlocal dielectric permittivity model. It was shown that a deviation from the *1/d$^2$* law was observed for separation distances in the angstrom range. In this case it is therefore clear that quantum effects will appear at larger distances than nonlocal effects. The goal of this paper is to pursue this work of Chapuis et al. [34] and to extend it to dielectric materials where *1/d$^2$* law occurs at much larger distances typically few hundreds of nanometers, in a domain where it is very likely to observea deviation from the local behavior at distances larger than quantum effects threshold distance.

As already suggested, we study in this paper the radiative heat transfer between two semi-infinite parallel dielectric solid planes as the gap distance $d$ between them tends to zero. We will carry on this study using a macroscopic nonlocal dielectric permittivity model suggested by Halevi and Fuchs [40] in which spatial dispersion is considered. The paper is organized as follows: in section II, we briefly review the near field radiative heat transfer calculation obtained in the framework of fluctuational electrodynamics formalism for a local modeling of the material optical response. In section III, we present the nonlocal modeling of the dielectric optical properties using the theory developed by Halevi and Fuchs. This theory is then used to calculate the radiative heat transfer coefficient between two 6H-SiC semi-infinite parallel planes. In section IV, we present the results obtained and discuss them comparing both local and nonlocal optical properties. Section V will be dedicated to the conclusions and future outlooks.

## II. Radiative Heat Transfer Formalism

Fluctuational electrodynamics introduced by Rytov [8,41] states that a body at a temperature $T$ radiates thermal energy due to the fluctuations of random currents generated by electrons in metals or ions in polar crystals. The properties of these currents are given by the fluctuation-dissipation theorem relating the currents correlation function (fluctuations) to the medium radiative losses (dissipation). These currents radiate an electromagnetic (EM) field related to the currents by the Green's tensors of the system. The emitted surfacedensity of the radiative heat flux (in W m$^{-2}$) is given by the Poynting vector $1/2\, Re[\langle \boldsymbol{E}(\boldsymbol{r},\omega) \times \boldsymbol{H}^*(\boldsymbol{r},\omega)\rangle]$, where $\boldsymbol{E}(\boldsymbol{r},\omega)$ and $\boldsymbol{H}(\boldsymbol{r},\omega)$ are the electric field and magnetic field, respectively.

In the most general sense, constitutive relations in a medium that relate bound charges to the electric field depend on the wavevector and the frequency so that for example $\boldsymbol{D}(\boldsymbol{k},\omega) = \epsilon(\boldsymbol{k},\omega)\boldsymbol{E}(\boldsymbol{k},\omega)$. When the EM field varies on a spatial scale larger than the microscopic characteristic lengths of the propagation medium, the medium is usually considered to be local so that $\boldsymbol{D}(\boldsymbol{r},\omega) = \epsilon(\boldsymbol{r},\omega)\boldsymbol{E}(\boldsymbol{r},\omega)$. When it is not the case, the medium is nonlocal i.e. the optical properties depend on the wavevector of the EM field [6,18].

As mentioned earlier, the surface density of the radiative heat flux $\phi$ between two semi-infinite parallel planes in local thermodynamic equilibrium, maintained at temperatures $T_1$ and $T_2$ and separated by a gap distance $d$ (Fig 1), can be calculated by means of fluctuational electrodynamics. When the temperature difference is small $(T_1-T_2)/T_1 \ll 1$, $\phi$ can be linearized and written as a radiative heat transfer coefficient (RHTC) $h$ multiplied by the temperature difference $\delta T$. The extended derivation of the RHTC has been done by many authors [2,3,6,9,42-46], and we just recall here the main expressions:

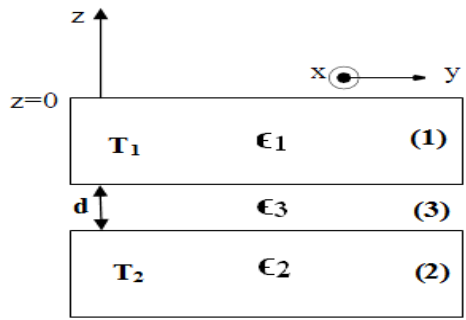

**Figure 1:** Two semi-infinite parallel material planes separated by a gap distance $d$.

$$\phi(T,d) = h_{rad}(T,d)\delta T \quad (1)$$

$$\begin{cases} h_{rad}(T,d) = \sum_{\alpha=S,P} \int_0^{+\infty} d\omega \left[ h_{prop}^{\alpha}(T,d,\omega) + h_{evan}^{\alpha}(T,d,\omega) \right] \\ h_{prop}(T,d,\omega) = h^0(T,\omega) \times \int_0^{k_0} \frac{KdK}{k_0^2} \frac{(1-|r_{31}^{\alpha}|^2)(1-|r_{32}^{\alpha}|^2)}{|1-r_{31}^{\alpha}r_{32}^{\alpha}e^{2i\gamma_3 d}|^2} \\ h_{evan}(T,d,\omega) = h^0(T,\omega) \times \int_{k_0}^{+\infty} \frac{KdK}{k_0^2} \frac{4Im(r_{31}^{\alpha})Im(r_{32}^{\alpha})e^{2i\gamma_3 d}}{|1-r_{31}^{\alpha}r_{32}^{\alpha}e^{2i\gamma_3 d}|^2} \end{cases} (2)$$

where $\omega$ is the wave angular frequency, $k_0 = \omega/c$, and $K$ and $\gamma_3 = \sqrt{\omega^2/c^2 - K^2}$ are the wavevector components parallel and normal to the surface in vacuum, respectively. It is worth mentioning here that due to the continuity conditions, $K$ is considered the same in all mediums. $r_{31}^{\alpha}$ and $r_{32}^{\alpha}$ represent the reflection factors for the EMwaves of polarization $\alpha = s, p$ incident from medium 3 and reflected on media 1 and 2, respectively. $h^0(T,\omega)$ is the derivative of the blackbody specific intensity of radiation with respect to temperature (Planck's law). These last quantities are given by the following equations:

$$\begin{cases} r_{3m}^p = \frac{\gamma_3 - \varepsilon_3 \omega Z_m^p}{\gamma_3 + \varepsilon_3 \omega Z_m^p} \\ r_{3m}^S = \frac{c^2 \gamma_3 Z_m^S - \omega}{c^2 \gamma_3 Z_m^S + \omega} \end{cases} (3)$$

$$h^0(T,\omega) = \frac{\hbar \omega^3}{4\pi^2 c^2} \frac{\hbar \omega}{k_B T^2} \left[ 2 \sinh\left(\frac{\hbar \omega}{2 k_B T}\right) \right]^{-2} (4)$$

Note that Eqs. (2) show that the RHTC is the sum of the contributions of propagative ($K < k_0$) and evanescent ($K > k_0$) waves of *s* and *p* polarizations. Note also that the reflection factors depend on the surface impedances $Z_m^{\alpha}$ between media 3 and *m* which are defined as the ratio of the parallel component of the electric field on the parallel component of the magnetic field.

Radiative heat transfer calculations were performed for 6H-type silicon carbide (SiC), a non-magnetic polar material characterized by a hexagonal crystallographic structure and a lattice constant ratio *c / a ≈ 4.9*. The crystallographic configuration of SiC is widely used in research and is studied especially at high temperatures due to its semiconducting and heat resistant properties [47].

Let us first recall what is happening in the local case. As an example, we consider two 6H-SiC semi-infinite parallel planes at temperatures $T_1 = 299.5\ K$ and $T_2 = 300.5\ K$ so that the average temperature of the system is $T = 300\ K$. We start by substituting the Lorentz-Drude local dielectric function given below in Eq. (5) [48], in the general equations of the surface impedances (seeEqs. (6) below)by assuming that the longitudinal and the transverse components of the dielectric function are equal in the static limit($\varepsilon(\omega) = \lim_{k \to 0} \varepsilon_t(k, \omega) = \lim_{k \to 0} \varepsilon_l(k, \omega)$).

$$\varepsilon(\omega) = \varepsilon_\infty \left( 1 + \frac{\omega_p^2}{\omega_T^2 - \omega^2 - i\gamma\omega} \right) \quad (5)$$

$$\begin{cases} Z_m^p = \dfrac{2i}{\pi\omega} \displaystyle\int_0^{+\infty} \dfrac{dq}{k^2} \left[ \dfrac{q^2}{\varepsilon_t(k,\omega) - (ck/\omega)^2} + \dfrac{K^2}{\varepsilon_l(k,\omega)} \right] \\ \\ Z_m^s = \dfrac{2i}{\pi\omega} \displaystyle\int_0^{+\infty} \dfrac{dq}{\varepsilon_t(k,\omega) - (ck/\omega)^2} \end{cases} \quad (6)$$

where $k^2 = q^2 + K^2$.

By substituting these equations into the expressions of the reflection factors as given by Eqs. (3), we obtain the classical Fresnel reflection factors:

$$\begin{cases} r_{3m}^p = \dfrac{\varepsilon_m \gamma_3 - \varepsilon_3 \gamma_m}{\varepsilon_m \gamma_3 + \varepsilon_3 \gamma_m} \\ r_{3m}^S = \dfrac{\gamma_3 - \gamma_m}{\gamma_3 + \gamma_m} \end{cases} \quad (7)$$

where $\gamma_m = \sqrt{\varepsilon_m (\omega/c)^2 - K^2}$. Then, we replaceEqs. (7) into the expression of the RHTC as given by Eqs. (2)to obtainits expression as a function of the separation distance $d$. We report in Fig 2, the calculated dependences of the different contributions to the RHTC.

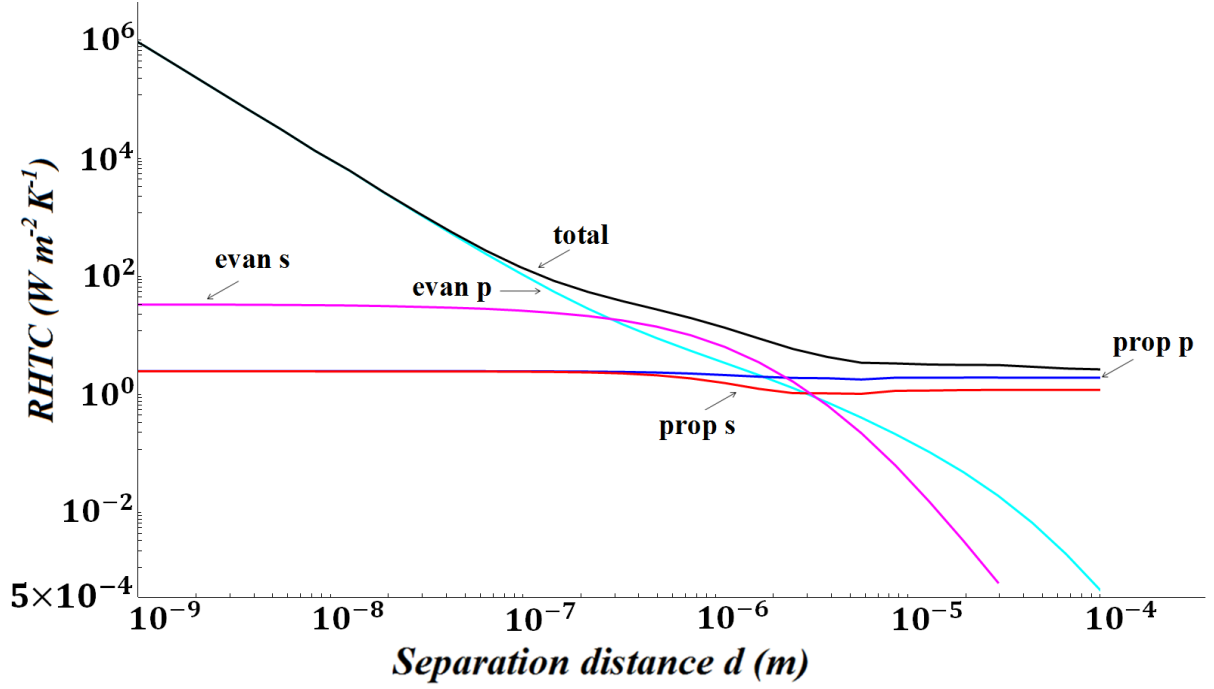

**Figure 2 :**Variation of the radiative heat transfer coefficient (RHTC) (evanescent and propagative contributions of EM waves of *s* and *p* polarizations) between two semi-infinite 6H-SiC parallel planes, for the local model case.

From these graphs, we observe that the evanescent EM wave *p* term has a well-known divergence behaving as $1/d^2$. This is due to the presence of surface polaritons on SiC which increases the density of EM states close to the surface as described in many articles [5-7,9,15,49-51]. In the case of *s* polarization, the RHTC saturates when the distance is smaller than the skin depth [34]. Note on the contrary that the contribution of propagative EM waves in both *s* and *p* polarizations does not change a lot for submicronic distances since the density of EM propagative states at small distances does not change significantly.

The divergence of the evanescent EM wave *p* polarization contribution cannot be physical at extremely small distances at which the EM fields begin to feel the microscopic variations of the matter properties. This led us, as few authors did before, to take into account the nonlocal behavior of matter by introducing a nonlocal dielectric permittivity function in order to overcome this problem.

## III. Nonlocal macroscopic dielectric permittivity function theory

Studying the nonlocal behavior of matter is not an easy task and is, to some extent controversial. The main problem is that in the presence of nonlocality, an incoming transverse EM wave gives birth not only to a single transverse wave in the material but also to a second transverse wave and a supplementary longitudinal wave. In this case, the usual boundary conditions on the continuity of the tangential components of **E** and **H** are not sufficient to solve the transmission problem of Maxwell's equations. Additional boundary conditions (ABC), often involving conditions on the polarization vector have to be set. However, in the literature, several ABC have been proposed [52-75].

Halevi and Fuchs [40] have suggested a theory in which all ABC (typically conditions on the component of polarization or its derivatives at the boundary) developed by different authors are included. The advantage of this theory is that it includes the main nonlocal modeling developed in the literature. Basically, spatial dispersion effects lead in the dielectric function expression to the addition of a term dependent on the square of the wavevector $k$. One of the simplest modeling is to use the single oscillator model in combination with the so-called hydrodynamic model [76]. The latter model has been used in a large variety of forms. In Halevi and Fuchs modeling, a spatial dispersion parameter $D$ is introduced. It is typically related to a diffusion phenomenon of the carriers in the medium. It is homogeneous to the square of a velocity divided by a typical frequency. Under these assumptions:

$$\varepsilon(\omega, k) = \varepsilon_\infty \left(1 + \frac{\omega_p^2}{\omega_T^2 - \omega^2 - i\nu\omega + Dk^2}\right) \quad (8)$$

where $D = \hbar\omega_T/(m_e + m_h)$, $\omega_T$ is the frequency of an isolated transition (for example an exciton), and $m_e$ and $m_h$ are the electron and hole masses, respectively. The frequency $\omega_P$ is a measure of the oscillator strength and $\nu$ represents the losses parameter. In the case of SiC, the parameters in Eq. (8) take the following values: $D = 1.77 \times 10^{10} m^2.s^{-2}$, $\omega_p = 1.049 \times 10^{14}\ rad.s^{-1}$, $\omega_T = 1.49 \times 10^{14}\ rad.s^{-1}$ and $\nu = 8.97 \times 10^{11}\ rad.s^{-1}$. As mentioned before, one has to add ABC in order to solve the reflection and transmission problems in Maxwell's equations. The ABC take the following forms as conditions on the polarization $\rho$ at the interface [40] which allows obtaining relations between the amplitudes of the waves (three transmitted waves and one reflected wave):

$$\alpha_i \rho_i(0^+) + \beta_i \partial \rho_i(0^+)/\partial z = 0 \quad i = x, z \quad (9)$$

where $\rho$ denotes the polarization. Eq. (9) apply for *p*-polarized EM wave. For *s*-polarized EM wave, one has a similar equation for $\rho_y(z)$.

**Table 1 :** Five different sets of surface scattering parameters proposed in literature.

| ABC | $U_x$ | $U_y$ | $U_z$ |
|---|---|---|---|
| Kliewer & Fuchs [52-56] | 1 | 1 | −1 |
| Rimbey & Mahan [57-61] | −1 | −1 | 1 |
| Pekar [62-65] | −1 | −1 | −1 |
| Ting et al. [54] | 1 | 1 | 1 |
| Agarwal et al. [66-75] | 0 | 0 | 0 |

These ABC therefore necessitate a choice of the ratio $\alpha_i/\beta_i$ of the parameters $\alpha_i$ and $\beta_i$. Different choices correspond to different surface scattering parameters (SSP), imposed in the expressions of the surface impedances and the reflection factors coefficients of the system. These parameters depend on the nature of the polarization of the EM field ($U_y$ for *s* polarization, $U_x$ and $U_z$ for *p* polarization). The derived expressions of the reflectivity and the susceptibility depend on these SSP. Halevi and Fuchs have made a correspondence between the SSP values ($U_x$, $U_y$ and $U_z$) and the ABC taken by different authors (Table 1).

$$\begin{cases} \dfrac{\alpha_j}{\beta_j} = i\Gamma \dfrac{1-U_j}{1+U_j} \qquad j = x, z \\ \Gamma = [(\omega^2 - \omega_T^2 + i\nu\omega - DK^2)/D]^{1/2} \end{cases} \quad (10)$$

We obtain the final expressions of the surface impedances by performing some algebra and introducing the parameters $a_l$ and $b_l$. The latter are given by:

$$\begin{cases} a_l = \dfrac{1}{q_l - \Gamma} + \dfrac{U_x}{q_l + \Gamma} \\ b_l = \left(\dfrac{1}{q_l - \Gamma} + \dfrac{U_z}{q_l + \Gamma}\right)\mu_l \quad l = 1,2,3 \\ \mu_1 = -\dfrac{K}{q_1}, \mu_2 = -\dfrac{K}{q_2}, \mu_3 = \dfrac{q_3}{K} \end{cases} \quad (11)$$

and the nonlocal surface impedances are expressed as:

$$\begin{cases} Z_p = \dfrac{(1,2)+(2,3)+(3,1)}{\varepsilon_1\left(\dfrac{k_0}{q_1}\right)(2,3)+\varepsilon_2\left(\dfrac{k_0}{q_2}\right)(3,1)}, (i,j)=a_i b_j - b_i a_j \\ Z_s = \dfrac{k_0(a_1-a_2)}{q_2 a_1 - q_1 a_2} \end{cases} \quad (12)$$

In Eq. (12) we made use of the definition $\varepsilon_l \equiv \varepsilon(q_l)$; $l=1,2,3$. The reflection factors at the surface are then obtained using the following general equations:

$$\begin{cases} r_p = \dfrac{Z_p - Z_p^{Local}}{Z_p + Z_p^{Local}} \\ r_s = \dfrac{Z_s - Z_s^{Local}}{Z_s + Z_s^{Local}} \end{cases} \quad (13)$$

where $Z_P^{Local}=q_1/\varepsilon_1 k_0$ and $Z_S^{Local}=k_0/q_1$.

Considering the nonlocal medium to be infinite, the frequency and the wavevector should satisfy the following dispersion equations for transverse and longitudinal waves, respectively:

$$\begin{cases} \varepsilon(\omega,k)=k^2/k_0^2 \\ \varepsilon(\omega,k)=0 \end{cases} \quad (14)$$

The solution of these equations gives three expressions for the $z$component of the wavevector ($q_z = q_1, q_2$ and $q_3$), that we substituted in the previous equations for each set of SSP to calculate thecorresponding surface impedances and reflection factors.

One can wonder what is the maximal spatial frequency for which the nonlocal modeling presented here remains valid. Clearly, in this modeling, the discrete nature of the atoms is not taken into account which will appear for typical sizes of the order of the atomic size i.e. in the angstrom range. This means that the modeling will lose its pertinence for spatial frequencies larger that $2\pi/10^{-10}$ or separation distances smaller than a fraction of a nanometer.

## IV. Results and discussions

Nonlocal RHTC variations as a function of the separation distance $d$ between two 6H-SiC semi-infinite parallel planes are plotted in Fig3. Each nonlocal graph corresponds to a different set of ABC. Up to a distance of approximately $d \approx 10^{-7}m$, the RHTC calculated in both local and nonlocal models are almost identical (see the inset). This is the domain of the local regime in the radiative heat transfer, where the use of a nonlocal dielectric function does not bring any change compared to the case where local EM properties are considered. Let us remind the reader what is happening in this regime:

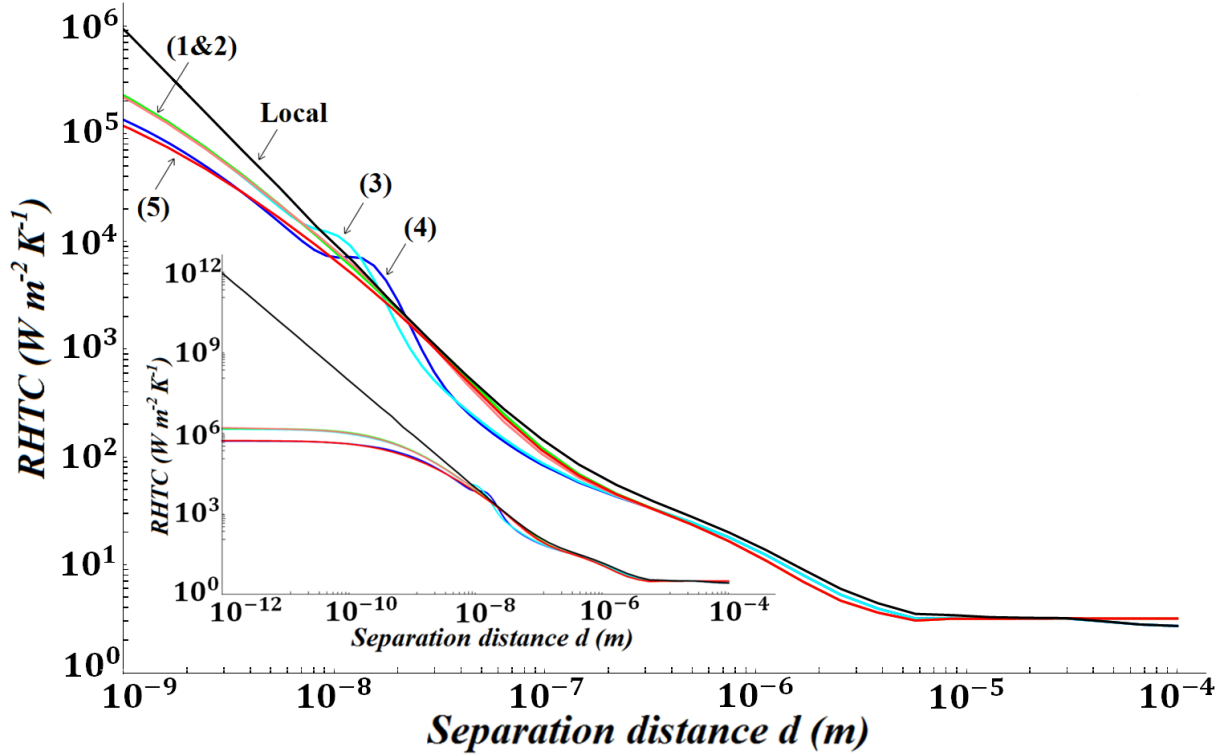

**Figure 3 :** Variation of the total radiative heat transfer coefficient (RHTC) between two semi-infinite 6H-SiC parallel planes, for the local model and the five ABC of the nonlocal model; (1): Rimbey & Mahan, (2): Agarwal et al., (3): Ting et al., (4): Kliewer and Fuchs, (5): Pekar. The inset shows the variation of the total RHTC for the nonlocal models in comparison with the local model.

At large distances compared to the thermal wavelength, exponentially decaying evanescent EM waves do not contribute to the RHTC. The value of the latter is then limited to the contribution of propagative EM waves and is somewhat less than the value $4\sigma T^3$. This is due to the fact that SiC is highly absorbent over a wide spectral range, except around $\lambda = 10.6$ µm where it is reflective. We also note that the term of the *p* polarized propagative EM waves gives values slightly higher than those of the *s* polarized propagative EM waves due to the existence of the Brewster angle for which the reflection contribution of the *p* polarized EM waves is zero and thus allowing greater absorption.

At subwavelength distances, some of the evanescent EM waves decay slowly (those with a small parallel wavevector *K* but *K* is still larger than $k_0$) so that these waves can tunnel between the surfaces. Their contributions to the RHTC can become dominant. Indeed, the contribution to the transfer (Eq. (2)) appears as a double integral over the angular frequency and the wave vector *K*. The integration domain in angular frequency is governed by the Planck spectrum emission band whereas the integration in *K* domain is typically between 0 and $2\pi/d$. For large wavevector *K*, the static limit of the reflection factor $r_{3m}^s = (\varepsilon - 1)/4(K/k_0)^2$ tends to zero leading to saturation of the heat flux at distances smaller than the skin depth[34].

Concerning the *p* polarized term, the reflection factor $r_{3m}^P \approx (\varepsilon - 1)/(\varepsilon + 1)$ gives a finite non-zero value for large $K$. If there is a frequency for which $\varepsilon = -1$, as it is the case for materials supporting surface waves [6], the contribution to the transfer will be very large at this frequency. As integration over $K$ is between 0 and $2\pi/d$, it easy to see from Eq. (2) that the transfer will follow a $1/d^2$ dependence dominated spectrally by the resonant frequency. Note, that this enhancement corresponds also to a large increase of the EM density of states which number at the surface increases as $1/d^2$ for small distances [6].

At distances of the order of $10^{-8} m$, one sees that the nonlocal graphs deviate from the $1/d^2$ asymptote. We note that this distance is of the order of the distance at which the term $Dk^2$ dominates in the denominator of the expression of the nonlocal dielectric function [Eq. (8)]. For sufficiently large $k$, the reflection coefficient will go to zero contrary to the local case. This means that the transfer is controlled by a critical wavevector limit and not by the inverse of the separation distance. Let us consider $k \sim 2\pi/d$ at a certain distance $d$ and the condition $Dk^2 \gg \omega_T^2$ in the denominator of Eq. (8), we find a critical distance $d \sim \sqrt{D 4\pi^2/\omega_T^2}$ approximately equals to $5 \times 10^{-9} m$ for which nonlocal behavior will be dominant. This distance can be seen as the distance travelled by the resonant heat carriers on an oscillation period at $\omega_T$. We therefore find that the nonlocal behavior occurs at distances of few nanometers, for which in principle quantum effects are still non-dominant since these effects have been reported at sub nanometer scale [35-38]. At distances of the order of 1 nm, the deviation of the nonlocal graphs from the local graph is significant and the values attained by these graphs are of one order of magnitude difference.

Moreover, in the graphs representing the nonlocal media with the ABC of Ting et al. and Kliewer and Fuchs, we note two bumps in the graphs at distances $d_1 \approx 1 \times 10^{-8} m$ and $d_2 = 2 \times 10^{-8} m$, respectively. It is not trivial to link these distances to the optical parameters. We have shown however by a parametric study that the bump position is closely related to the $\omega_p$ value and is almost insensitive to the value of the losses parameter *v* in Eq. (8).

At subnanometric separation distances, all radiative heat transfer calculation obtained with different ABC have very similar behaviors. They all saturate to a certain value that can be considered as the ultimate radiative conductance between two semi-infinite parallel planes of 6H-SiC. Note that ultimately small values of the separation distance ($10^{-12} m$) taken in the inset graph of Fig. 3 are nonphysical but they are considered just in order to show that the nonlocal matter description mathematically leads to a saturation value in the radiative heat transfer. This conductance is around $10^6$ $W$ $m^{-2}$ $K^{-1}$. Note that this conductance is much smaller than the one which is obtained in conduction if we make the ratio of the thermal conductivity of SiC ($\kappa$=400 W m$^{-1}$ K$^{-1}$) on the size of the typical distance between atoms in SiC($r_o = 15.1 \times 10^{-10} m$).This means that heat transfer by radiation is always beaten by conduction heat transfer in the matter. This also means that when the distances are going to reduce as small as 1 nm, other effects such as quantum effects, that are completely different from electromagnetic effects described here have to be taken into account to describe the full heat transfer process. As this work limits itself to radiative heat transfer, this quantum treatment is beyond the scope of this paper.

The saturation value of the thermal radiation can also be interpreted in terms of the number of coupled modes. Heat transfer can actually be written in the Landauer way as a summation over the system eigenmodes of the product of the number of modes by the mean energy carried by each mode and by the transmission coefficient of the mode through the cavity. Each mode of the system being determined by the angular frequency and the parallel wavevector, summation is performed over these two quantities. The transmission coefficients are given by the following equations for the propagative and evanescent contributions, respectively:

$$\begin{cases} h_{rad}(T,d) = \sum_{\alpha=S,P} \int_0^{+\infty} d\omega h^0(T,\omega) \int_0^\infty \frac{KdK}{k_0^2} \tau(\omega,K) \\ \tau(\omega, K < k_0 = \omega/c) = \frac{(1-|r_{31}^\alpha|^2)(1-|r_{32}^\alpha|^2)}{|1-r_{31}^\alpha r_{32}^\alpha e^{2i\gamma_3 d}|^2} \\ \tau(\omega, K > k_0) = \frac{4Im(r_{31}^\alpha)Im(r_{32}^\alpha)e^{2i\gamma_3 d}}{|1-r_{31}^\alpha r_{32}^\alpha e^{2i\gamma_3 d}|^2} \end{cases} (15)$$

Finally, after integration over $\omega$ and $K$, the RHTC can be seen as the total number of coupled modes per surface unit multiplied by the quantum of the thermal conductance $g_0 = \pi^2 k_b^2 T/3h$ which can be seen as the rate at which heat is transported by a bosonic carrier channel. Therefore the number of modes per surface unit at 300 K can be estimated and is around 3 x10$^{15}$ coupled modes per m$^2$.

In order to understand which modes contribute to the radiative heat transfer when the two SiC surfaces are approached one to each other, we plot the transmission coefficient for the evanescent EM waves $4(Im(r_{31}^P))^2 e^{2i\gamma_3 d}/|1-(r_{31}^P)^2 e^{2i\gamma_3 d}|^2$ at different separation distances for the local model and the nonlocal model with Kliewer and FuchsABC. The transmission coefficient plots in the $(\omega, K)$ plane represented as a function of the angular frequency $\omega$ and the parallel wavevector *K* are reported in Fig. 4. We note that for a separation distance of 100 nm, the transmission coefficients in both cases are very similar. The modes are very well coupled ($\tau = 1$) for the modes corresponding to coupled surface phonon-polaritons of SiC in the cavity. For the local dielectric modeling case, the transmission coefficient map has a similar shape when the separation distance is reduced except that more and more modes contribute to the transfer. We see that the same map shape is obtained as long as we increase the parallel wavevector scale as the inverse of the separation distance. This explains why the transfer increases as $1/d^2$ and why the spectral contributions to the transfer are always occurring at the same frequencies. Indeed, as the separation distance decreases, the shape of the transfer spectrum does not change except that the scale increases as $1/d^2$. This spectrum is narrow and the transfer occurs around surface-polaritons frequencies.

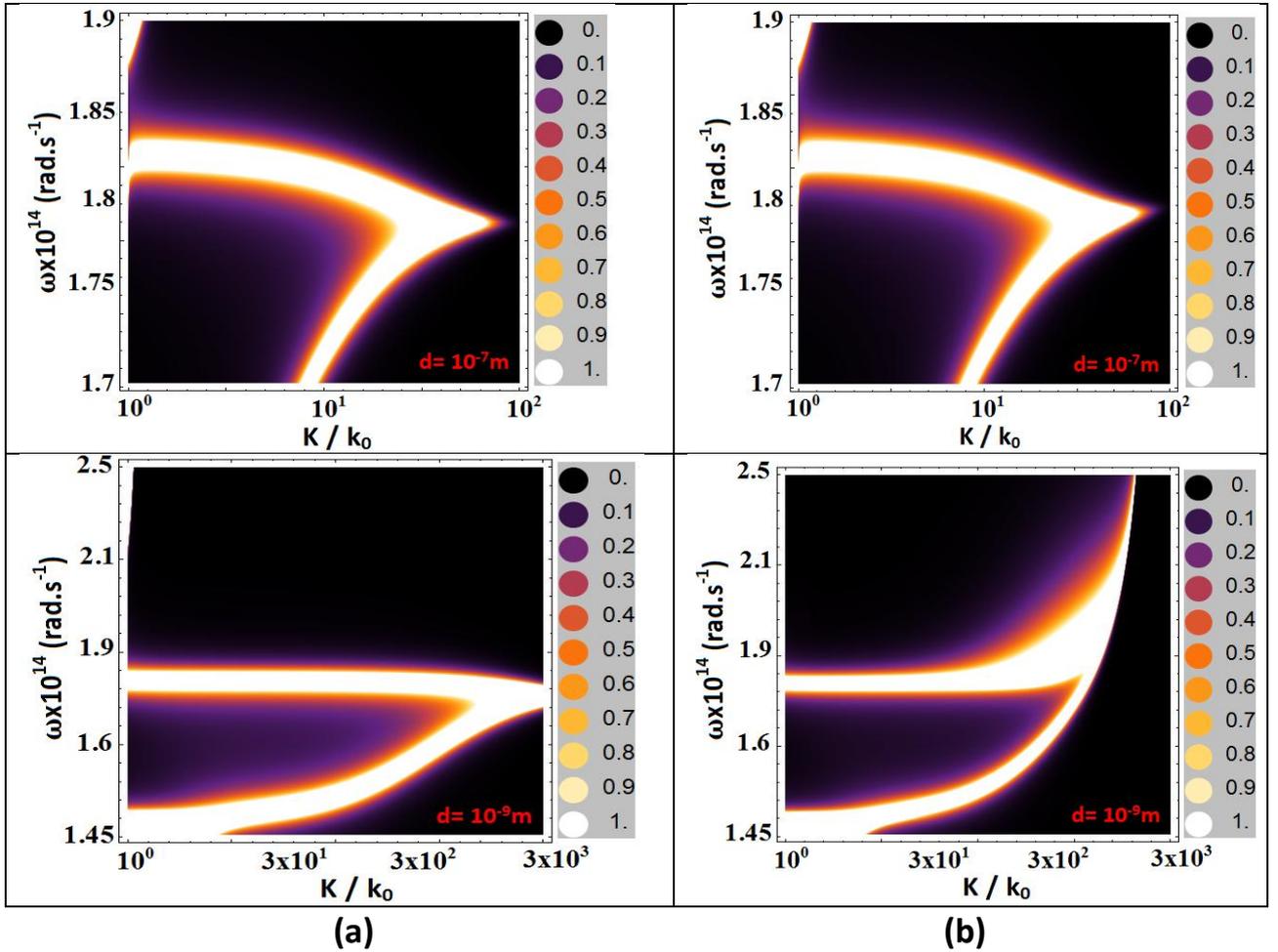

**Figure 4:** 2D-plot in the plane (ω,K) of the transmission coefficient $4(Im(r_{31}^P))^2 e^{2i\gamma_3 d}/|1-(r_{31}^P)^2 e^{2i\gamma_3 d}|^2$ of the p-polarization evanescent EM waves for the local case (a) and the nonlocal case of Kliewer and Fuchs ABC (b) at different separation distances $d$.

On the other hand, the case of the nonlocal modeling of the dielectric function, shows a somewhat different situation. We note that most of the transfer still occurs around phonon-polariton angular frequencies. However, by decreasing the distance, the transmission coefficient map starts to show a clear cut-off in the parallel wavevector. Contrary to the local case, for separation distances below 1 nm, the transmission coefficient map does not change. We note that the angular frequency domain at which the transfer occurs broadens. Moreover, there are no modes able to well couple for parallel wavevector larger than few hundreds of $k_0$. This can be seen in Fig. 5 where the radiative transfer spectrum (a) is represented with the density of EM energy spectrum (b). We see that the spectrum broadens and saturates as the distance is reduced. We also show that the transfer spectrum is very similar to the energy density spectrum. This is not surprising since this last quantity is directly proportional to the local density of EM states (LDOS) and the transfer spectrum is also related to the LDOS.

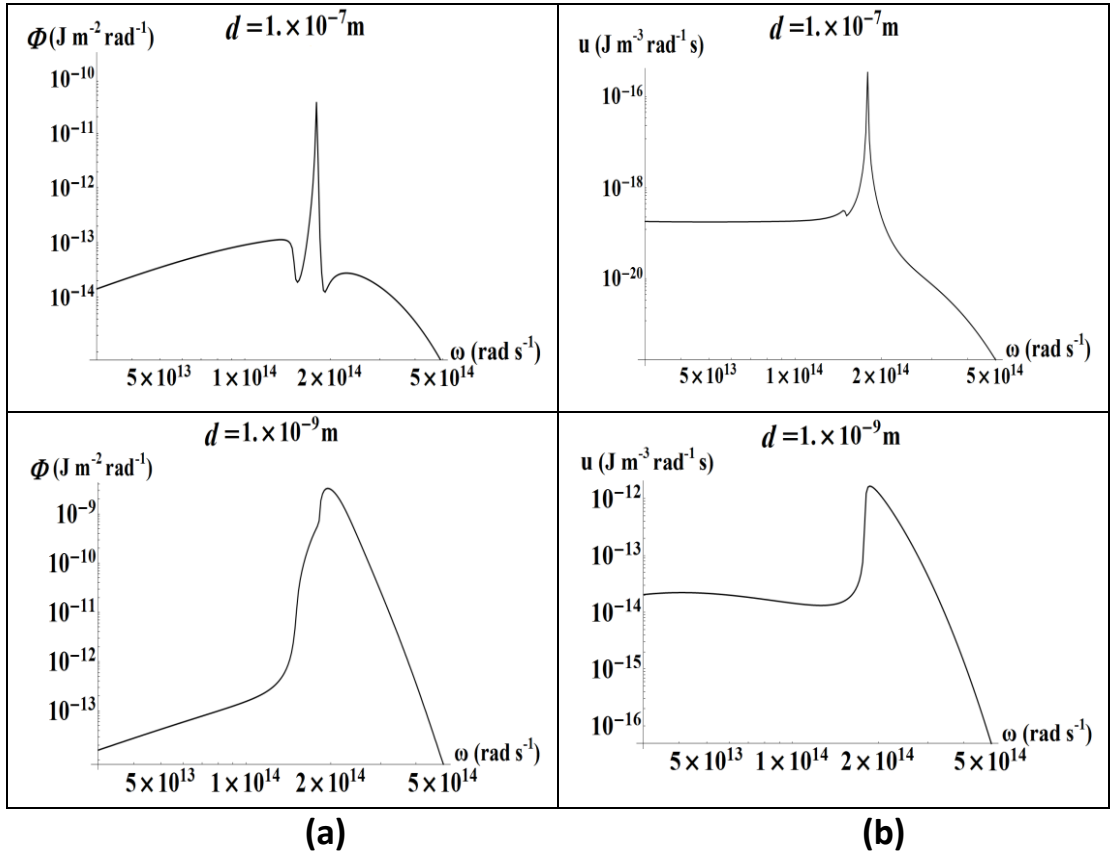

**Figure 5:** Plots of the spectral energy flux (a) and the spectral EM energy density (b) of the P-polarization evanescent EM waves as functions of the angular frequency for the nonlocal case of Kliewer and Fuchs ABC at different distances $d$.

# V. Conclusions

We have studied in this work the radiative heat transfer between two semi-infiniteparallel dielectric 6H-SiC planes taking into account the nonlocal corrections in the material optical properties. We chose to followHalevi and Fuchs nonlocal dielectric permittivity function theory that considers scattering of the electromagnetic excitation at the surface of the dielectric material and which includes most of the different nonlocal modeling of dielectrics. This assumption leads to defineadditional boundary conditions (ABC) needed to solve the transmission problem in Maxwell's equations. These ABC appear as additional surface scattering parameters in the derived expressions of the surface impedances and reflection factors. Taking into account the spatial dispersion that is given as an additional term depending on the square of the total wavevector in the dielectric permittivity function, we studied the above mentioned different cases to calculate the radiative heat transfer coefficient (RHTC). We showed that for separation distances between few nanometers and few hundreds of nanometers, the RHTC follows a $1/d^2$ dependence law identical for both nonlocal and local material optical responses. On the other hand, at distances of few nanometers, the RHTC calculated with nonlocal modeling deviates from $1/d^2$ law: heat transfer is also broadened when compared to the local case.

Different features were revealed from the RHTC graphs, as two bumps appeared for the cases of Kliewer andFuchsand Ting et al. ABC.Saturation of the flux in the nonlocal case is obtained for distances much smaller than the atomic size, where the modeling presented here more likely ceasesto be valid. At sub nanometer scale, heat transfer by electromagnetic waves probably ceases to be the dominant transfer process and quantum effects enter into play leading to a transition between radiation and conduction [35-38].

In futureworks,we will have to compareour theoretical resultswith experimentalmeasurements of near field thermal radiation. This would allow us to determine at the same time the distance at which the radiative heat transfer stops to bethe dominant heat transfer process (below 1nm) as well as the distance where local medium approximation becomes not valid (few nanometers). Experiment measurement could also be a wayto choosebetween the different ABC that are suggested in the literature. The existence or non-existence of "bumps" could eliminate some of the modeling approachesandsuggest a consistent nonlocaldielectricpermittivity function model for dielectrics.